\documentclass[11pt]{article}
\usepackage[T1]{fontenc}
\usepackage[latin1]{inputenc}
\usepackage[dvips]{epsfig}
\usepackage{graphicx}
\usepackage[english]{babel}
\usepackage{amsmath}
\usepackage{amssymb}
\usepackage{amsfonts}
\title{\bf $f(T)$ cosmology via Noether symmetry}
\author{K. Atazadeh$^{1}$\thanks{email: atazadeh@azaruniv.edu}
 \,and F. Darabi$^{1,2}$\thanks{email:
f.darabi@azaruniv.edu}
\\ {\small $^1$Department of Physics, Azarbaijan
University of Tarbiat Moallem, Tabriz 53741-161, Iran}\\$^{2}${\small Research Institute for Astronomy and Astrophysics of Maragha (RIAAM), Maragha 55134-441, Iran}}

\begin{document}
\maketitle
\begin{abstract}
We consider Noether symmetry approach to find out exact cosmological solutions in $f(T)$-gravity. Instead of taking into account phenomenological models, we apply the Noether symmetry to the $f(T)$ gravity.  As a result, the presence of such symmetries selects viable models and allow to solve  the equations of motion. We
show that the generated $f(T)$ leads to a power law expansion for the cosmological scale factor.
\end{abstract}
\vspace{2cm}

\section{Introduction}
The expansion of the universe is currently undergoing a period of acceleration which is directly
measured from observations such as
Type Ia Supernovae~\cite{SN1}, \cite{SN2}
cosmic microwave background (CMB) radiation~\cite{WMAP, Komatsu:2010fb},
large scale structure~\cite{LSS},
baryon acoustic oscillations~\cite{BAO},
and weak lensing~\cite{WL}.
There are two remarkable
approaches to explain the late time acceleration
of the universe:
One is to introduce some unknown matters called ``dark energy''
in the framework of general relativity
(for a review on dark energy, see, e.g.,~\cite{Copeland:2006wr, Li:2011sd}).
The other is to modify the gravitational theory,
e.g., $f(R)$ gravity~\cite{
DeFelice:2010aj, Clifton:2011jh}.

To extend gravity beyond general relativity,
``teleparallelism" could be considered by using
the Weitzenb\"{o}ck connection, which has no curvature but
torsion, rather than the curvature defined by the Levi-Civita connection
\cite{Weitzenb}, \cite{Hehl:1976kj}.
This approach was also taken by Einstein~\cite{Einstein}.
The teleparallel Lagrangian density described by the torsion scalar $T$
has been promoted to a function of $T$, $i.e.$, $f(T)$,
in order to account for the late time cosmic
acceleration~\cite{Bengochea:2008gz}
as well as inflation~\cite{Inflation-F-F}.
This concept is similar to the idea of $f(R)$ gravity.
Various aspects of
$f(T)$ gravity have been examined in the
literature~\cite{f(T)-Refs,f(T)-Refs1,f(T)-Refs2,f(T)-Refs3,
Local-Lorentz-invariance}.
In particular, the presence of extra degrees of freedom and the violation of local Lorentz invariance as well
as the existence of non-trivial frames for $f(T)$ gravity have been noted~\cite{Local-Lorentz-invariance}.
Evidently, more studies on $f(T)$ gravity are needed to see if the theory is a viable one.
For a comprehensive review of the teleparallel gravity, the reader is referred to \cite{Andrade}.

In this paper, we consider a flat FRW space-time
in the framework of the metric formalism of $f(T)$ gravity.
Following \cite{7}, we set up an effective Lagrangian in which the
scale factor $a$ and torsion scalar $T$ play the role of independent
dynamical variables. This Lagrangian is constructed in such a way
that its variation with respect to $a$ and $T$ yields the correct
equations of motion as that of an action with a generic $f(T)$
mentioned above. The form of the function $f(T)$ appearing in the
modified action is then found by demanding that the Lagrangian
admits the desired Noether symmetry \cite{8,10}. For a study of the Noether symmetry in $f(R)$ cosmology see \cite{9}. Similarly, we shall see that by
demanding the Noether symmetry as a feature of the Lagrangian of
the cosmological model under consideration, we can obtain the explicit form of
the function $f(T)$. Since the existence of a symmetry results in
a constant of motion, we can integrate the field equations which
would then lead to a power law expansion of the universe.

\section{ $f(T)$ gravity and cosmology}
\label{sec2}

To consider teleparallelism, one employs the orthonormal tetrad components
$e_A (x^{\mu})$, where an index $A$ runs over $0, 1, 2, 3$ to the
tangent space at each point $x^{\mu}$ of the manifold.
Their relation to the metric $g_{\mu\nu}$ is given by
\begin{equation}
g_{\mu\nu}=\eta_{A B} e^A_\mu e^B_\nu\,,
\label{eq:2.1}
\end{equation}
where $\mu$ and $\nu$ are coordinate indices on the manifold
and also run over $0, 1, 2, 3$,
and $e_A^\mu$ forms the tangent vector on the tangent space over which the
metric $\eta_{A B}$ is defined.

Instead of using the torsionless Levi-Civita connection in General Relativity,
we use the curvatureless Weitzenb\"{o}ck connection in Teleparallelism
\cite{Weitzenb}, whose non-null torsion $T^\rho_{\verb| |\mu\nu}$ and contorsion
$K^{\mu\nu}_{\verb|  |\rho}$ are defined by
\begin{eqnarray}
T^\rho_{\verb| |\mu\nu} \equiv e^\rho_A
\left( \partial_\mu e^A_\nu - \partial_\nu e^A_\mu \right)\,,
\label{eq:2.2}
\end{eqnarray}
\begin{eqnarray}
K^{\mu\nu}_{\verb|  |\rho}
\equiv
-\frac{1}{2}
\left(T^{\mu\nu}_{\verb|  |\rho} - T^{\nu \mu}_{\verb|  |\rho} -
T_\rho^{\verb| |\mu\nu}\right)\,,
\label{eq:2.3}
\end{eqnarray}
respectively.
Moreover, instead of the Ricci scalar $R$ for the Lagrangian density
in general relativity, the teleparallel Lagrangian density is described by the torsion scalar $T$ as follows
\begin{equation}
T \equiv S_\rho^{\verb| |\mu\nu} T^\rho_{\verb| |\mu\nu}\,,
\label{eq:2.4}
\end{equation}
where
\begin{equation}
S_\rho^{\verb| |\mu\nu} \equiv \frac{1}{2}
\left(K^{\mu\nu}_{\verb|  |\rho}+\delta^\mu_\rho \
T^{\alpha \nu}_{\verb|  |\alpha}-\delta^\nu_\rho \
T^{\alpha \mu}_{\verb|  |\alpha}\right)\,.
\label{eq:2.5}
\end{equation}

The modified teleparallel action for $f(T)$ gravity
is given by
\begin{equation}
I=
\int d^4x |e|  f(T)\,,
\label{eq:2.6}
\end{equation}
where $|e|= \det \left(e^A_\mu \right)=\sqrt{-g}$ and
the units has been chosen so that $c=16\pi G=1$. Note that in action (\ref{eq:2.6}), we have overlooked any matter contribution in the action.
Varying the action in Eq.~(\ref{eq:2.6}) with respect to
the vierbein vector field $e_A^\mu$, we obtain the equation~\cite{Bengochea:2008gz}
\begin{equation}
\frac{1}{e} \partial_\mu \left( eS_A^{\verb| |\mu\nu} \right) f_{T}
-e_A^\lambda T^\rho_{\verb| |\mu \lambda} S_\rho^{\verb| |\nu\mu}
f_{T} +S_A^{\verb| |\mu\nu} \partial_\mu \left(T\right) f_{_{TT}}
+\frac{1}{4} e_A^\nu f = 0\,,
\label{eq:2.7}
\end{equation}
where a subscript $T$ denotes differentiation with respect to $T$.
We assume the four-dimensional flat
Friedmann-Lema\^{i}tre-Robertson-Walker (FLRW)
space-time with the metric,
\begin{equation}
d s^2 = h_{\alpha \beta} d x^{\alpha} d x^{\beta}
+\tilde{r}^2 d \Omega^2\,,
\label{eq:2.8}
\end{equation}
where $\tilde{r}=a(t)r$, $x^0=t$ and $x^1=r$ with the two-dimensional
metric $h_{\alpha \beta}={\rm diag}(-1, a^2(t))$.
Here, $a(t)$ is the scale factor and
$d \Omega^2$ is the metric of two-dimensional sphere with unit radius.
In this space-time,
$g_{\mu \nu}= \mathrm{diag} (-1, a^2, a^2, a^2)$ and
the tetrad components $e^A_\mu = (1,a,a,a)$
yield the exact value of torsion scalar
\begin{equation}\label{torsion}
T=-6H^2,
\end{equation}
where $H=\dot{a}/a$ is the Hubble parameter
and the dot denotes the time derivative of $\partial/\partial t$.

In the flat FLRW background,
it follows from Eq.~(\ref{eq:2.7}) that
the modified Friedmann equations are given
by~\cite{Bengochea:2008gz}
\begin{eqnarray}\label{fri}
12f_{_{T}}H^2+f=0\,,
\label{eq:2.9}
\end{eqnarray}
\begin{eqnarray}
\dot{H}
=\frac{1}{ 4T\,f_{_{TT}} + 2f_{_{T}} }
\left(-T\,f_{_{T}}+\frac{f}{2} \right)\,.
\label{eq:2.10}
\end{eqnarray}

It is known that $f(T)$ gravity has first-order gravitational field equation in derivatives. Similar to $f(R)$ gravity in general relativity
where the gravitational field equation is fourth-order in derivatives,
it is important to investigate the theoretical aspects in order to examine whether $f(T)$ gravity can be a gravitational theory like $f(R)$ gravity.

In order to derive the cosmological equations in a FLRW metric, one
can define a canonical Lagrangian ${\cal L}={\cal L}(a,\dot{a}, T,
\dot{T})$, where  ${\cal Q}=\{a,T\}$ is the configuration space
and ${\cal TQ}=\{a,\dot{a}, T, \dot{T}\}$ is the related tangent
bundle on which ${\cal L}$ is defined. The variable $a(t)$ and
$T(t)$ are the scale factor and the  torsion scalar in the FLRW
metric, respectively. One can use the method of the Lagrange
multipliers to set  $T$ as a constraint of the dynamics. Selecting
the suitable Lagrange multiplier and integrating by parts, the
Lagrangian ${\cal L}$ becomes canonical. In our case, we have

\begin{equation}
I=2\pi^2\int dt\,a^3 \left\{f(T)-\lambda\left[T+6\left(\frac{\dot a^2}{a^2}\right)\right]
\right\} ,
\end{equation}
where
$a$ is the scale factor scaled with respect to today's value (so
that $a=\tilde a/\tilde a_0$ and $a(t_0)=1$). This choice for $a$, makes it
dimensionless, whereas $[f]=M^4$. It is straightforward to show that,
for $f(T)=-T$, one obtains the usual
Friedmann equations.

The variation with respect to $T$ of the action gives
$\lambda=f_T$. Therefore the previous action can be rewritten as
\begin{equation}
I=2\pi^2\int dt\,a^3\left\{f-f_T\left[T+6\left(\frac{\dot a^2}{a^2}\right)\right]
\right\}\, ,
\end{equation}
 and then, integrating by parts, the point-like FLRW Lagrangian is obtained
\begin{equation}
{\cal L}= a^3\,(f-f_T\,T)
-6\,f_T\,a\,\dot a^2
,\label{eqz0}
\end{equation}
which is a canonical function of two coupled fields, $T$ and $a$, both depending on time $t$. The momenta conjugate to variables $a$ and $T$ are
\begin{equation}\label{pa}
p_{a}=\frac{\partial{\cal L}}{\partial\dot{a}}=-12f_{T}a\dot{a}.
\end{equation}
\begin{equation}
p_{T}=\frac{\partial{\cal L}}{\partial\dot{T}}=0.
\end{equation}
The total energy $E_{\cal L}$ corresponding to the $0,0$-Einstein equation is
\begin{equation}
\label{energy}
E_{{\cal L}}=-
6\,f_T\,a\,\dot a^2- a^3\,(f-f_T\,T)=0.
\end{equation}
The equations of motion for $T$ and $a$ are
\begin{equation}
\label{T}
a^{3}f_{_{TT}}\left(T+6\frac{\dot{a}^2}{a^2}\right)=0,
\end{equation}
 \begin{equation}
\label{a}
-6f_T\,H^2-12f_T\frac{\ddot{a}}{a}=3(f-f_T\,T)+12f_{_{TT}}\,\dot{T}\,H,
\end{equation}
respectively. Note that from the equations (\ref{energy}), (\ref{T}) and (\ref{a}) we can recover the equation (\ref{eq:2.9}), (\ref{torsion}) and (\ref{eq:2.10}), respectively. Here we have taken $f_{TT}\neq0$.

Considering $T$ and $a$ as the variables,  we see that $T$
coincides with the definition of the torsion scalar in the FLRW
metric (excluding the case $f_{_{TT}}=0$).  Geometrically, this is the Euler constraint of the dynamics.
Furthermore, as we will show below, constraints on the form of the
function $f(T)$ can be achieved by asking for the existence of
Noether symmetries. On the other hand, the existence of
the Noether symmetries guarantees the reduction of dynamics and
the eventual solvability of the system.

\section{The Noether symmetry}

Solutions for the dynamics given by the point-like Lagrangian (\ref{eqz0}) can be obtained by selecting cyclic variables related to some Noether symmetry \cite{9,99}. In principle, as a physical criterion, this approach
allows one to select $f(T)$ gravity models which are compatible with the Noether symmetry.
Let $\mathcal{L}(q^i,\dot{q}^i)$ be a canonical, non degenerate point-like Lagrangian so that
\begin{equation}
\frac{\partial\mathcal{L}}{\partial U}=0, \hspace{1.5cm}  det H_{ij}\equiv \left\|\frac{\partial^2\mathcal{L}}{\partial \dot{q}^i\partial \dot{q}^j}\right\|\neq0,
\end{equation}
where $H_{ij}$ is the Hessian matrix related to the Lagrangian $\mathcal{L}$ and a dot denotes derivative with respect to the affine parameter $U$, namely
the cosmic time $\textit{t}$. The Lagrangian in analytic mechanics is of the form
\begin{equation}
\mathcal{L}=T(\textbf{q},\dot{\textbf{q}})-V(\textbf{q}),                   \label{4.39}
\end{equation}
where \textit{T} and \textit{V} are the positive definite quadratic `kinetic energy' and `potential energy', respectively. The energy function associated with $\mathcal{L}$ is:
\begin{equation}
E_\mathcal{L}\equiv\frac{\partial\mathcal{L}}{\partial \dot{q}^i}-\mathcal{L},
\end{equation}
which is the total energy $T + V$ and the constant of motion. Since our cosmological problem has a finite number of degrees of freedom, we are going to consider only point transformations.  Any invertible transformation of the `generalized positions' $Q^i=Q^i(\textbf{q})$ induces a a local transformation of the `generalized velocities' such that:
\begin{equation}
\dot{Q}^i(\textbf{q})=\frac{\partial Q^i}{\partial q^j}\dot{q}^j;   \label{4.23}
\end{equation}
the matrix $\mathcal{J}=\left\|\partial Q^i/\partial q^j\right\|$ is the Jacobian of the transformation on the positions, and it is assumed to be non-zero. A point transformation $Q^i=Q^i(\textbf{q})$ can depend on a (or more than one) parameter. As starting point, we can assume that a point transformation depends on a parameter $\epsilon$, so that $Q^i=Q^i(\textbf{q},\epsilon)$, and that it gives rise to a one-parameter Lie group. For infinitesimal values of $\epsilon$, the transformation is then generated by a vector field: for instance, $\partial/\partial x$ is a translation along the $x$ axis while $x(\partial/\partial y)-y(\partial/\partial x)$ is a rotation around the $z$ axis and so on.

In general, an infinitesimal point transformation is represented by a generic vector field on $Q$
\begin{equation}
\textbf{X}=\alpha^i(\textbf{q})\frac{\partial}{\partial q^i}.
\end{equation}
\\ The induced transformation (\ref{4.23}) is then represented by
\begin{equation}
\textbf{X}^c=\alpha^i\frac{\partial}{\partial q^i}+\left(\frac{d}{d\lambda}\alpha^i(\textbf{q})\right)\frac{\partial}{\partial \dot{q}^j}.   \label{4.24}
\end{equation}
A function $F(\textbf{q}, \dot{\textbf{q}})$ is invariant under the transformation \textbf{X} if
\begin{equation}
L_XF\equiv\alpha^i(\textbf{q})\frac{\partial F}{\partial q^i}+\left(\frac{d}{d\lambda}\alpha^i(\textbf{q})\right)\frac{\partial}{\partial \dot{q}^j}F=0,
\end{equation}
where $L_XF$ is the Lie derivative of \textit{F}. If $L_X\mathcal{L}=0$, \textbf{X} is then a symmetry for the dynamics derived by $\mathcal{L}$.
Let us now consider a Lagrangian $\mathcal{L}$ and its Euler-Lagrange equations
\begin{equation}
\frac{d}{d\lambda}\frac{\partial\mathcal{L}}{\partial \dot{q}^j}-\frac{\partial\mathcal{L}}{\partial q^j}=0.                \label{4.25}
\end{equation}
Contracting (\ref{4.25}) with $\alpha^i$s leads to
\begin{equation}
\alpha^j\left(\frac{d}{d\lambda}\frac{\partial\mathcal{L}}{\partial \dot{q}^j}-\frac{\partial\mathcal{L}}{\partial q^j}\right)=0.     \label{4.26}
\end{equation}
Using the total derivative relation as
\begin{equation}
\alpha^j\frac{d}{d\lambda}\frac{\partial\mathcal{L}}{\partial \dot{q}^j}=\frac{d}{d\lambda}\left(\alpha^j\frac{\partial\mathcal{L}}{\partial \dot{q}^j}\right)-\left(\frac{d\alpha^j}{d\lambda}\right)\frac{\partial\mathcal{L}}{\partial \dot{q}^j},
\end{equation}
we obtain from equation (\ref{4.26}) that
\begin{equation}
\frac{d}{d\lambda}\left(\alpha^j\frac{\partial\mathcal{L}}{\partial \dot{q}^j}\right)=L_X\mathcal{L}.
\end{equation}
The immediate consequence is the \textit{Noether theorem} which states:
if $L_X\mathcal{L}=0$, then the function
\begin{equation}
\Sigma_0=\alpha^k\frac{\partial\mathcal{L}}{\partial \dot{q}^k},                                                \label{4.27}
\end{equation}
is a constant of motion.
Equation (\ref{4.27}) can be expressed independently of coordinates as a contraction of \textbf{X} by a Cartan 1-form
\begin{equation}
\theta_\mathcal{L}\equiv\frac{\partial\mathcal{L}}{\partial \dot{q}^j}dq^j.
\end{equation}
For a generic vector field $\textbf{Y}=y^i\partial/\partial x^i$, and 1-form $\beta=\beta_idx^i$, we have by definition $i_Y\beta=y^i\beta_i$. Thus equation (\ref{4.27}) can be expressed as:
\begin{equation}
i_X\theta_\mathcal{L}=\Sigma_0.
\end{equation}
Through a point transformation, the vector field \textbf{X} becomes:
\begin{equation}
\tilde{\textbf{X}}=\left(i_XdQ^k\right)\frac{\partial}{\partial Q^k}+\left(\frac{d}{d\lambda}\left(i_XdQ^k\right)\right)\frac{\partial}{\partial \dot{Q}_k}.
\end{equation}
If $\textbf{X}$ is a symmetry and we choose a point transformations such that
\begin{equation}
i_XdQ^1=1; \hspace{1.2cm}   i_XdQ^i=0 \hspace{1cm}  i\neq1,                           \label{4.38}
\end{equation}
we get
\begin{equation}
\tilde{\textbf{X}}=\frac{\partial}{\partial Q^1};  \hspace{1.2cm}   \frac{\partial \mathcal{L}}{\partial Q^1}=0.
\end{equation}
Consequently, $Q^1$ is a cyclic coordinate, related to conserved quantities, reducing the dynamics of the system to a manageable one. Furthermore, the change of coordinates given by (\ref{4.38}) is not unique. In general, the solution of equation (\ref{4.38}) is not defined on the whole space, rather, it is local in the sense explained above.

Following \cite{8,10}, we define the Noether symmetry induced on the model by a vector field $X$ on the tangent space $T{\cal Q}=\left(a,T,\dot{a},\dot{T}\right)$ of the configuration space ${\cal Q}=\left(a,T\right)$ of Lagrangian
(\ref{eqz0})
\begin{equation}\label{J}
X=\alpha \frac{\partial}{\partial a}+\beta
\frac{\partial}{\partial T}+\frac{d
\alpha}{dt}\frac{\partial}{\partial \dot{a}}+\frac{d
\beta}{dt}\frac{\partial}{\partial \dot{T}},
\end{equation}
such that the Lie derivative of the Lagrangian with respect to
this vector field vanishes
\begin{equation}\label{K}
L_X {\cal L}=0.
\end{equation}
In (\ref{J}), $\alpha$ and $\beta$ are functions of $a$ and $T$
and $\frac{d}{dt}$ represents the Lie derivative along the
dynamical vector field, that is,
\begin{equation}\label{L}
\frac{d}{dt}=\dot{a}\frac{\partial}{\partial
a}+\dot{T}\frac{\partial}{\partial T}.
\end{equation}
It is easy to find the constants of motion corresponding to such a
symmetry. Indeed, equation (\ref{K}) can be rewritten as
\begin{equation}\label{M}
L_X{\cal L}=\left(\alpha \frac{\partial {\cal L}}{\partial
a}+\frac{d\alpha}{dt}\frac{\partial {\cal L}}{\partial
\dot{a}}\right)+\left(\beta \frac{\partial {\cal L}}{\partial
T}+\frac{d\beta}{dt}\frac{\partial {\cal L}}{\partial
\dot{T}}\right)=0.
\end{equation}
Noting that $\frac{\partial {\cal
L}}{\partial q}=\frac{dp_q}{dt}$, we have
\begin{equation}\label{N}
\left(\alpha\frac{dp_a}{dt}+\frac{d\alpha}{dt}p_a\right)+\left(\beta\frac{dp_T}{dt}+\frac{d\beta}{dt}p_T\right)=0,
\end{equation}
which yields
\begin{equation}\label{O}
\frac{d}{dt}\left(\alpha p_a+\beta p_T\right)=0.
\end{equation}
Thus, the constants of motion are
\begin{equation}\label{P}
Q=\alpha p_a+\beta p_T,
\end{equation}
whereas in $f(T)$ theory we have $p_{T}=0$ and so the corresponding constant of motion becomes $Q=\alpha p_{a}$.

In order to obtain the functions $\alpha$ and $\beta$ we use
equation (\ref{M}). In general, this equation gives a
polynomial in terms of $\dot{a}^2$ and $\dot{a}\dot{T}$ with coefficients
being partial derivatives of $\alpha$ and $\beta$ with respect to
the configuration variables $a$ and $T$. Thus, the resulting
expression is identically equal to zero if and only if these
coefficients are zero. This leads to a system of partial
differential equations for $\alpha$ and $\beta$.
\section{Noether symmetries in $f(T)$ cosmology}\label{sec4}
For the existence of a symmetry, we can write the following system of equations (linear in $\alpha$ and $\beta$),
\begin{equation}\label{eqz1}
f_T(\alpha+2a\,\partial_{_a}\alpha)+a\,f_{TT}\beta=0,
\end{equation}
\begin{equation}\label{eqz2}
a\,f_T\,\partial_{_T}\alpha=0.
\end{equation}
which are obtained by setting the coefficients of the terms
$\dot{a}^2$ and $\dot{a}\dot{T}$ in $L_{\bf X}{\cal
L}=0$ to zero. In order to make $L_{\bf X}{\cal L}=0$ vanish we will also
look for those particular $f$'s which, given the Euler dynamics,
also satisfy the constraint
\begin{equation}\label{eqz4}
3\alpha\,(f-T\,f_T)-a\,\beta\,T\,f_{TT}=0\, .
\end{equation}
This procedure is different from the usual Noether symmetry
approach, in the sense that now $L_{\bf X}{\cal L}=0$ will be
solved not for all dynamics (which solve the Euler-Lagrange
equations), but only for those $f$ which allows Euler solutions to
solve also the constraint (\ref{eqz4}). Imposing such a constraint
on the form of $f$ will turn out to be, as we will show, a sufficient condition to find solutions of the Euler-Lagrange
equation which also possess a constant of motion, i.e.\ a Noether
charge. As we shall see later, the system (\ref{eqz1}) and
(\ref{eqz2})  can be solved exactly. Having a
non-trivial solution for $\alpha$ and $\beta$ for this system, one
finds a constant of motion if also the constraint (\ref{eqz4}) is
satisfied.
A solution of (\ref{eqz1}) and (\ref{eqz2}) exists
if explicit forms of $\alpha$, $\beta$ are found. If, at least one
of them is different from zero, a Noether symmetry exists.

When $f_{T}\neq0$, from equation (\ref{eqz2}) we get
\begin{equation}\label{1}
 \partial_{_T}\alpha=0\, \Rightarrow \alpha=\alpha(a).
 \end{equation}
Thus, from this equation it can be seen $\alpha$ only depend on $a$.
On the other hand, we can obtain $\alpha$ from equation (\ref{eqz4}) as follows
\begin{equation}\label{2}
\alpha(a)=\frac{a\,f_{_{TT}}T}{3(f-T\,f_{_{T}})}\beta(a,T),
\end{equation}
If one uses this expression in equation (\ref{eqz1}) to eliminate
$\alpha(a)$, one obtains
\begin{equation}\label{AD}
\frac{\partial \beta}{\partial a}=\frac{-3f}{2a\,T\,f_{_{T}}}\beta(a,T).
\end{equation}
To solve this equation we assume that the function $\beta (a,T)$
can be written in the form $\beta (a,T)=A(a)B(T)$, where $A$ and
$B$ are separate functions of $a$ and $T$, respectively. Substituting
this ansatz for $\beta (a,T)$ into equation (\ref{AD}), we obtain
\begin{equation}\label{AE}
\frac{2a}{A}\frac{dA}{da}=-\frac{3f}{T\,f_{_{T}}}.
\end{equation}
Since the left-hand side of this equation is a function of $a$
only while the right-hand side is a function of $T$, we should
have
\begin{equation}\label{AF}
-\frac{3f}{T\,f_{_{T}}}=C=\mbox{Const},
\end{equation}
which results in
\begin{equation}\label{AG}
f(T)=f_{_{0}}T^{(-\frac{3}{C})}.
\end{equation}
On the other hand equation (\ref{AE}), with its right hand-side equal
to $C$, has the solution
\begin{equation}\label{AH}
\frac{2a}{A}\frac{dA}{da}=C \Rightarrow
A(a)=a^{C/2}.
\end{equation}
Now, using (\ref{AG}) and $\beta(a,T)=a^{C/2}B(T)$ in equation
(\ref{2}), we find
\begin{equation}\label{AI}
\alpha(a)=\frac{1}{C}T^{-1}\,a^{\frac{C}{2}+1}B(T).
\end{equation}
Since $\alpha(a)$ should be a function of $a$ only, from the above
expression for $\alpha(a)$ we can write
\begin{equation}
B(T)=T,
\end{equation}
and thus we obtain
\begin{equation}\label{AJ}
\beta(a,T)=T\,a^{C/2},\hspace{.5cm}\alpha(a)=\frac{1}{C}a^{\frac{C}{2}+1}.
\end{equation}

To obtain the corresponding cosmology resulting from this type of
$f(T)$, we note that the existence of Noether symmetry
implies the existence of a constant of motion $Q=\alpha p_a$. Hence, using equation (\ref{pa})  we have
\begin{equation}\label{AO}
Q=36\frac{a^{(\frac{C}{2}+2)}\dot{a}\,T^{-(1+\frac{3}{C})}}{C}.
\end{equation}
On the other hand, by taking $f_{_{TT}}\neq0$  from equation (\ref{T}) we obtain
\begin{equation}\label{AR}
T=-6\frac{\dot{a}^{2}}{a^{2}}.
\end{equation}
By using equations (\ref{AO}) and (\ref{AR}) we can obtain  scale factor as following
\begin{equation}\label{AU}
a(t) \sim t^{\frac{-2}{C}}.
\end{equation}
Therefore, in the context of $f(T)=f_{_{0}}T^{-(\frac{3}{C})}$
cosmology, the universe evolves with a power law expansion. It
is seen from (\ref{AU}) that the condition under which the
universe expands is $C<0$.
The deceleration parameter  as a function of $C$ is therefore given by
\begin{equation}
q(C)=-\left(1+\frac{C}{2}\right)
\end{equation}
The condition for acceleration is $q(C) < 0$, thus we have $C>-2$. As it can be seen, for $C\rightarrow0$ we have $q \rightarrow-1$, that is the universe finally approaches the eternal de Sitter phase with infinite
acceleration, however, the accelerated expansion occurs in $-2<C<0$.

\section{Dark energy equation of state and age of the Universe in $f(T)$ }
In this section, we consider
the accelerated expansion of the universe in the context of $f(T)$ theory, without introducing  the mysteries fluid the so-called ``dark energy'' with a negative equation of state (EOS) parameter $w$. Let us then start from
equation (\ref{AU}) and find the Hubble parameter as a function of the redshift $z$ as
\begin{equation}\label{z}
H(z)=-\frac{2}{C}H_{0}(1+z)^{-\frac{C}{2}},
\end{equation}
where $a_{0}/a = 1+z$ with $a_{0}$ and $H_{0}$ being the values of the parameter at the present
epoch. It is worth noting that equation (\ref{z}) is the same as that derived from the standard
Friedmann equation with $w = -2/3$. Now, we may write the Friedmann equation
in a formal fashion which would encapsulate any modification to the standard
Friedmann equation in the last term regardless of its nature \cite{linder} that is
\begin{equation}\label{hh}
H^{2}/H^{2}_{0} = \Omega_{m}(1 + z)^{3} + \delta H^{2}/H^{2}_{0} ,
\end{equation}
where $\Omega_{m} = \rho/\rho_{0c}$, $\rho_{0c} = 3H^{2}_{0}$. Also, defining the effective EOS, denoted by
$w_{_{\rm eff}}(z)$, as

\begin{equation}\label{hhh}
w_{_{\rm eff}}(z) = -1+\frac{1}{3}\frac{d \ln\delta H^{2}}{d \ln(1 + z)} ,
\end{equation}
we can calculate $w_{_{\rm eff}}(z)$ using equations (\ref{z}), (\ref{hh}) and (\ref{hhh}) with the result
\begin{equation}\label{kk}
w_{_{\rm eff}}=-1+\frac{1}{3}\frac{-\frac{1}{C}(1+z)^{-c}-3\Omega_{m}(1+z)^{3}}{\frac{4}{C^{2}}(1+z)^{-c}-\Omega_{m}(1+z)^{3}}.
\end{equation}
Figure 1 shows the behavior of the effective EOS parameter,
$w_{_{\rm eff}}$, as a function of $z$. As it can be seen, for $C=-1$ ($-2<C\leq-1$) and $\Omega_{m}=0.33$  we have $w_{_{\rm eff}}\lessapprox-1$, which is the characteristic of one type of dark energy,
the so-called phantom and from equation (\ref{kk}), for $C\rightarrow0^{-}$,  we have $w_{_{\rm eff}}\rightarrow-1$.
\begin{figure}
\begin{center}
\epsfig{figure=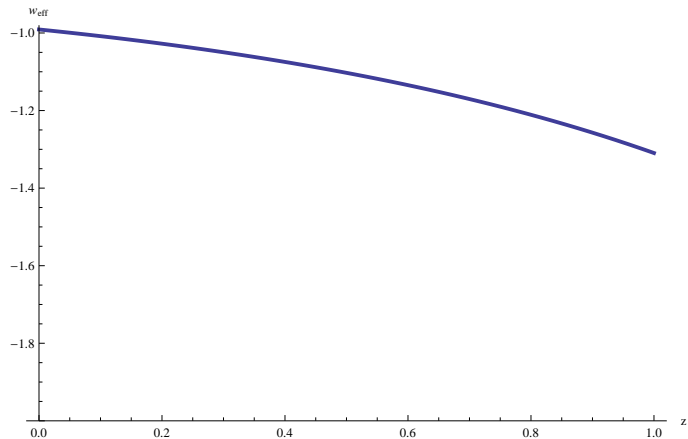,width=9cm}
\end{center}
\caption{\footnotesize The behavior of effective EOS,  $w_{_{\rm eff}}$, as a function of $z$. When $C=-1$ and
$\Omega_{m}=0.33$, we get a phantom behavior in the model.}
\end{figure}

To continue we consider the age of the Universe in $f(T)$ model. Thus, the age of the matter dominated Universe in FLRW models is given by
\begin{equation}
H_{0}t_{0}=\frac{2}{3}\frac{1}{\sqrt{1-\Omega_{m}}}\ln\left[\frac{1+\sqrt{1-\Omega_{m}}}{\sqrt{\Omega_{m}}}\right]
\end{equation}
where $H_{0}^{-1} = 9.8 × 10^{9}h^{-1}$ years and the dimensionless parameter $h$, according to resent data, is
about $0.7$. Hence, in the flat matter dominated universe with $\Omega_{total} = 1$ the age of the universe would
be only 9.3 Gyr, whereas the oldest globular clusters yield an age of about $\sim13.5$  Gyr \cite{27}. We obtain the age of the universe for our model from equations (\ref{fri}), (\ref{AG}) and (\ref{AU}) by taking matter in the Friedmann equations as follows
\begin{equation}
H_{0}t_{0}=\frac{-2}{C}\frac{1}{\sqrt{1-\Omega_{m}}}\ln\left[\frac{1+\sqrt{1-\Omega_{m}}}{\sqrt{\Omega_{m}}}\right]
\end{equation}
For a flat,  matter dominated Universe with $\Omega_{m} \approx 0.3$ and $C =-3$ we have a prediction for the
age of the Universe of about $13.2$ Gyr. It seems that the age of the universe in our model is longer
than the FLRW model. Figure 2 shows the behavior of the dimensionless age parameter, $H_{0}t_{0}$, as a function
of $\Omega _{m}$ for different values of $C$.
\begin{figure}
\begin{center}
\epsfig{figure=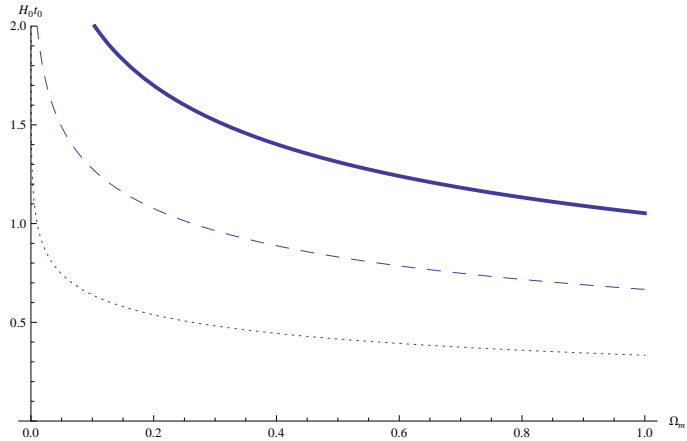,width=9cm}
\end{center}
\caption{\footnotesize $H_{0}t_{0}$ as a function of
$\Omega_{m}$ for $C = -1.9$ (solid line), $C =-3$ (dashed line) and $C = -6$ (dotted line). Figure shows that for a fixed value of $\Omega_{m}$ the predicted age of the universe is longer for larger values of $C$.}
\end{figure}
\\
\\
\begin{center}
{\bf Note added}
\end{center}
Before sending our present work to the Archive (on 13 December 2011), we became aware of the fact that there already existed a paper
arXiv:1112.2270 [gr-qc]\cite{Hao} (sent to the Archive on 10 December 2011), in which the authors had addressed similar issues. Although the content
and results of the present work overlap with the ones in \cite{Hao}, it is worth noting that our work was completed independently.

\section*{Acknowledgment}
This work has been supported financially by Research Institute
for Astronomy and Astrophysics of Maragha (RIAAM) under research project
No. 1/2361.

\section{Conclusions}
In this paper we have studied a generic $f(T)$ cosmological model
by using Noether symmetry approach. We have taken the background geometry as a flat FLRW metric and derived the
general equations of motion in this background. The phase space has been constructed by taking the scale factor $a$ and torsion scalar
$T$ as two independent dynamical variables. The Lagrangian of the
model in the configuration space spanned by $\left\{a,T\right\}$ is
so constructed that its variation with respect to these
dynamical variables yields the correct field equations. The
existence of Noether symmetry implies that the Lie derivative of
this Lagrangian with respect to the infinitesimal generator of the
desired symmetry vanishes. By applying this condition to the
Lagrangian of the model, we have obtained the explicit form of the
corresponding $f(T)$ function. We have shown that this form of
$f(T)$ results in a power law expansion for the scale factor and the accelerated expansion occurs in $-2<C<0$

\end{document}